# Artificial Intelligence in the Legal Field: Law Students' Perspective

Daniela Andreeva[¶,1] and Guergana Savova[±], PhD
[¶]London Metropolitan University, London, UK
[±]Computational Health Informatics Program, Boston Children's Hospital and Harvard Medical School, Boston, MA

October 13, 2024

## Abstract

The Artificial Intelligence (AI) field experienced a renaissance in the last few years across various fields such as law, medicine, and finance. Firms have been devoting significant resources to developing and evaluating AI applications including the creation of internal AI Innovation Departments. While there are studies outlining the landscape of AI in the legal field as well as surveys of the current AI efforts of law firms, to our knowledge there has not been an investigation of the intersection of law students and AI. Such research is critical to help ensure current law students are positioned to fully exploit this technology as they embark on their legal careers but to also assist existing legal firms to better leverage their AI skillset both operationally and in helping to formulate future legal frameworks for regulating this technology across many industries. The study presented in this paper addresses this gap – the intersection of law students and AI. Through a survey of 27 questions conducted over the period of July 22 – Aug 19, 2024, the study covers the law students' background, AI usage, AI applications in the legal field, AI regulations and open-ended comments to share opinions. The results from this study show the uniqueness of law students as a distinct cohort. The results differ from the ones of established law firms especially in AI engagement -- established legal professionals are more engaged than law students. The law firm participants show much higher enthusiasm about AI than this student cohort -- somewhat surprising as one would expect the younger generation to be more open to new technologies. Collaborations with Computer Science departments would further enhance the AI knowledge and experience of law students in AI technologies such as prompt engineering, chain-of-thought prompting, zero- and few-shot prompting, and language model hallucination management. As future work, we would like to expand the study to include more variables and a larger cohort more evenly distributed across locales. In addition, it would be insightful to repeat the study with the current cohort in one year's time to track how the students' viewpoints evolve.

## Introduction

Since the late 2010's the field of Artificial Intelligence (AI) has assertively re-emerged in the public space due to the convergence of plentiful digital data, the advances in hardware designs and capabilities, and scientific inventions in the representation of unstructured data including text, images, video, speech, and music. The scientific inventions of representations such as the transformer[1] and its flavors, operate on massive amounts of unstructured data to learn language, facts and pixels enabled by unprecedented computational hardware that performs mind-boggling matrix calculations. While the definition of AI and especially Artificial General Intelligence (AGI) remains fluid in the current discourse of both academic and commercial communities (for a review see Gebru and Torres[2]), from a practical point of view and for the purposes of this paper, we adopt the operational definition of AI as a set of computational tools for processing the unstructured data – mainly text and the rise of large language models (generalist LLMs[2], also referred to as foundation models) – to assist the legal professional in their daily work-related activities. One might argue that these tools are predominantly from the field of Natural Language Processing (NLP) – the field that endeavors to devise computational methods for transforming, understanding and generating language -- because they deal with classic NLP tasks such as information extraction, summarization, text generation, and question-answering. Although the NLP field is related to AI, it is not absorbed by it; however, we leave these nuances aside to focus on our investigation into the law students' views, opinions and concerns about AI as the students prepare for and grow into their careers.

---

[1] Corresponding author. LLB student at London Metropolitan University.
Email: danielazandreeva97@gmail.com

A recent position paper by Kapoor et al[3] performs a careful analysis of the promises and pitfalls of AI for legal applications and underscores its most important use case – information processing, e.g. summarization, translation, e-discovery. Other applications discussed are referred to as creativity, reasoning and judgment (e.g. drafting, automated mediation and dispute resolution) and prediction (e.g. predicting court decision). The paper emphasizes the importance of the involvement of legal professionals in the evaluation process of language models to understand how the technology can best aid those in the field, a thesis also raised by others (e.g. Cheong et al[4], Ma et al[5] Bhambhoria et al[6]). Most importantly, these studies highlight cases where AI should not be used such as arguing court cases, decision making, and judgments. These stances align with the AI TESCREAL philosophical framework developed by Gebru and Torres where TESCREAL denotes transhumanism, Extropianism, singularitarianism, (modern) cosmism, Rationalism, Effective Altruism, and longtermism[3]. Gebru and Torres urge AI practitioners to work on tasks *"for which we can develop safety protocols, rather than attempting to build a presumably all-knowing system such as AGI"*.

A 2024 white paper by Ma et al overviews the generative AI Legal landscape[5]. Generative AI is a subclass of LLMs where the transformer method focuses on the preceding sequence to generate the next target, e.g. the popular ChatGPT[7] family of models. As legal knowledge is expressed in all types of unstructured form -- text, images, videos, audio -- the relevance of these technologies is poignant. The authors review emerging technical solutions such as retrieval-augmented generation (RAG)[8], prompt engineering[9], fine-tuning, alignment, and the importance of legal datasets to train LLMs. They offer a legal technology market analysis across five main segments of legal work: (1) research and analysis, (2) document processing/contracting, (3) litigation, (4) time and billing, and (5) legal operations. The authors estimate the legal technology market at ~13B USD in 2023 growing at 3.5% compound annual growth rate. A separate study by Bloomberg lists the best current legal AI Tool options in the legal tech market[10]. The Ma et al white paper raises the structural concern that *"perhaps underdiscussed, is the notion that the general workforce, particularly those who are young lawyers entering the profession, may be ill prepared for the types of work that will result from the automation of legal tasks by generative AI."*

---

[2] LLMs refer to Large Language Models; not to be confused with LL.M. which stands for Master of Laws degree. We use both LLM and LL.M. to refer to the relevant intended meaning. LLB or LL.B stands for Legum Baccalaureus, Bachelor of Laws.

[3] Following Gebru and Torres "*we have capitalized some of the TESCREAL ideologies but not others. We are following what members of these ideologies themselves capitalize: Extropians, Rationalists, and Effective Altruists prefer to capitalize these terms, whereas transhumanists, singularitarians, cosmists, and longtermists do not*".

A study by Mahari – a law student -- investigates the opinions and action items regarding AI undertaken by law firms[11]. The study includes 59 senior and managing law firm partners researching their views on and engagement with AI. It shows that legal practitioners have embraced AI in the legal field and highlights the need for further practice-oriented research in collaboration with NLP experts. Over 95% of the respondents state that their company is engaged with AI in some form (discussions, conferences, training, etc.). Over 90% of the participants express a positive attitude towards AI in the legal field (cautiously hopeful to excited). Only a small percentage of the participants have little engagement with AI or a concerned attitude towards the technology.

In parallel to these developments, the law student community has been increasingly active in demonstrating their interest in AI applications of their future professional field. The UK-based law firm Slaughter and May has been organizing Student Innovation competitions since 2020[12] drawing about 70 entries each year. The 2024 competition focused on generative AI with the winning entry exploring the use of generative AI in Mergers and Acquisitions (M&A) practice[13]. Other listed entries dealt with topics such as "*What are the cultural challenges with integrating generative AI into a law firm*", "*Pick a practice area of your choice (litigation was picked) and explain a potential use case for generative AI, setting out how and why it will add value to the firm and its clients.*" A blog post by a law student[14] describes that "*because of concerns about the accuracy and authenticity of generative AI only 28% of students would use tools like ChatGPT to help with their applications*[4]*.*". Prof. David Wilkins from Harvard Law School raises the question of how to best prepare students for the new realities of legal careers[15]. There are an increasing number of blogs and websites on the topics of law digital transformation geared to include early stage legal professionals and academic viewpoints[16]. Topics cover technology developments as well as policy implications[17] and regulatory frameworks, e.g. the EU AI Act[17].

Our study builds on the previous work and focuses on the intersection of AI and the future generation of legal professionals, i.e. current law students. It contributes to the growing body of investigations into AI in the legal field as it aims to take the pulse of the views, opinions, awareness and uses of AI by law students in the legal field and potential areas of improvement – complexity, accessibility, reliability, regulation and privacy being at the top. The paper is organized as follows. We describe our methods, then present the results. The Discussion section provides salient points for consideration. It is important to note that the attitude shared by the authors of this paper is one of a cautious but non-limiting approach to the future and AI in the legal field and far from the all-encompassing utopic views as defined – and critiqued -- in the TESCREAL bundle[2].

---

[4] Applications refer to the vacation scheme or training contract application forms and materials.

## Methods

We created a survey comprising of a list of 27 questions, split into 5 sections – Background (Q1-3), AI Usage (Q4-12), AI Applications in the Legal Field (Q13-20), AI Regulations (Q21-25), and Conclusion (Q26-27). The topics of the questions are listed in Table 1 in the Appendix. We used Google Forms[18] to create the survey. We pilot tested the survey with three volunteers before we distributed it broadly.

The survey aimed at a wide demographic of law students studying at a range of different locations, universities and levels of studies. It was open for the period of July 22 – Aug 19, 2024.

We posted the survey on LinkedIn and in relevant LinkedIn groups. It reposted frequently to ensure visibility. In addition, the survey was distributed via a student WhatsApp group as well as direct outreach to law students. Personalized messages were sent to each participant, explaining the aim of the study and providing details. Continued communication was established with the participants to ensure clarity and engagement through which we ascertained their student status. However, we did not formally vet the participants' student status.

Participation in the study was voluntary and the data were securely stored within the internal systems of our research team. The results were aggregated; therefore, confidentiality of the participants was maintained.

The data were cleaned, structured and analyzed using Microsoft Excel[19]. Tables and figures were created with the aggregated data to ensure the readability of the results.

## Results

Table 1 in the Appendix lists the question topics from the survey and the breakdown per type of answer. The survey attracted attention from undergraduate, graduate students and early stage working professionals. There are 54 participants.

Survey Section -- Background (Q1-3) ▮ : The participants are predominantly undergraduate law students (LLB students) (63%), 1st year (36%) followed by 3rd year (21%) and LL.M (26%) and from the UK. The distribution shows participants at different levels, across the UK and some outside of the UK (e.g. Nigeria, Pakistan, India).

Survey Section -- AI Usage (Q4-12) ▮ : Most of the participants state that they do not have any AI training (91%), although they feel they have solid (54%) or good/user (43%) understanding of AI applications but not necessarily the knowledge of the technical underpinnings. This understanding is gained via the duration of their AI usage -- 75% indicate usage of more than a month but less than two years, 10% had 2+ years of usage. Of the participants who use AI applications, many of them use them multiple times per day (28%), with 18% of the participants indicating weekly usage of more than 5 times. The preferred tool is the free ChatGPT 3.5 version[20] (65%) although some of them use both the free and the subscription versions of ChatGPT[21] (20%). No participant states usage of open LLMs, e.g. LLaMA 2/3[22] which is perhaps due to the lack of more user-friendly interfaces for these models which in turn requires some level of computer science skills.

Furthermore, the participants actively use the AI tools to help with both or either school and work (73%). They also use them for other tasks, e.g. planning and scheduling, financial investments, cooking recipes, writing emails. Overall, they find AI helpful (98% spread across "extremely helpful" 37%, "helpful" 27% and "somewhat helpful" 34%) and they do not indicate any bias or prejudice in using it.

Survey Section -- AI Applications in the Legal Field (Q13-20) 🟩 : Although the majority of the participants find that AI applications are important in the Legal field (74%) and that they are excited about them (50% spread across 20% excited and 30% mostly hopeful), only a small portion of all participants (20%) indicated use of any of the more law-specific platforms. The platforms the participants listed include Aria[23], Chatsonic[24], Harvey[25], ContractPodAi[26], Luminance[27] and ChatGPT[7].

3% of the participants state that their university offers a formal training in AI in the Legal field as part of the curriculum. The vast majority (86%) consider such training important. In addition, a large portion (74%) are enthusiastic about collaborations with Computer Science departments on applications of AI in the Legal field. The participants expressed the feeling that AI skills are needed in the legal profession now (36%) and definitely in five years (51%) when most of the survey participants will enter the workforce.

We left Q20 open-ended to allow recording of consequential applications of AI in the Legal field. The participants were very active with their suggestions and enumerate 25 use cases. Among them, the top ones are Drafting (especially contracts but also pleadings, judgments, general statistics, and emails), summarization, legal research, information extraction and data science (collection, extraction, tracking, data-driven decision making). Some participants felt that all aspects of the Legal field could benefit from AI applications. Other use cases include case findings, interpretation of judgments and precedents, judicial administration, marriage/deeds of assignment, topic understanding, and proofreading.

Survey Section -- AI Regulations (21-25) 🟨 : 38% of the participants feel that AI regulations do not adequately keep up with the developments in the AI technology field, while 23% feel otherwise. Most of the participants follow the EU AI Act[17] and the activities of the UK Office of Artificial Intelligence[28]. However, only a small group (19%) follows current legal cases involving AI, e.g. *New York Times v OpenAI*[29].

Survey Section -- Conclusion (Q26-27) 🟪 : Q26 aimed to gauge participants' interest in participating in an AI and Commercial Awareness Law Student Club as a venue to keep up-to-date of technology and legal developments in the area. The majority of the students are interested in such a club. Q27 is open-ended asking to share any ideas or thoughts on the topic of AI in the Legal Field. Twelve respondents shared opinions which we include in the Discussion section.

## Discussion

The results of this study demonstrate the uniqueness of law students as a distinct cohort when researching AI in the legal field and the need for in-depth investigations to understand its awareness and concerns. The Mahari study on generative AI in law firms included in the Introduction section[11] show that 95% of the law firms are engaged with AI; compare that with 67% of our student participants indicating they feel excellently, well or somewhat prepared for future usage of AI in the legal field. The law firm participants in the Mahari study[11] show much higher level of enthusiasm about AI in the legal field compared to our student participants (90% vs. 50%). This difference might be attributed to the level of experience in and exposure to relevant AI tools between the participants in the two studies (senior/managing law firm partners vs. law students). The two studies align in the expressed desire for interdisciplinary collaborations between the legal and AI/computer science communities with Mahari listing concrete initiatives that might be undertaken[11].

Our results and the differences with the Mahari study[11] point to the desire on the part of the students for AI training. Although there are excellent LL.M.(see footnote 2 for abbreviation) programs[30] focused on AI, they appear to be mainly at the Master's level. Law student AI training could be a collaborative effort with Computer Science departments which are well suited to teach the technical underpinnings of the technologies including the intricacies of robust tool evaluations and technology limitations (bottom-up perspective). Of course, such courses could take many shapes[31], perhaps with minimal exposure to coding as it could be intimidating for non-computational students and focus more on concepts and actual applications, e.g. prompt engineering[9], chain-of-thought prompting[32], zero- and few-shot prompting[9], LLM hallucination management[33] in the context of legal tools. Exposure to the AI philosophical, regulatory and policy frameworks would enhance early-stage legal professionals' thinking of positioning AI in the broader context (top-down perspective). The convergence of the top-down and bottom-up perspective of viewing AI would contribute to a cohesive vision. The learning outcomes of such training would prepare the early-stage legal professionals not only for the use of AI in the workspace and for leadership positions in the Innovation departments within their firms, but also would prepare the next generation of jurists who will regulate AI as well as represent and adjudicate AI-related cases.

Furthermore, none of the student cohort participants indicates experience with open models such as Llama family of models[22] which have consistently shown performance results on common evaluation benchmarks competitive with closed models such as ChatGPT. The use of open models usually requires some level of programming skills. Therefore, as a call to open model creators is to make available easy-to-use interfaces to stimulate their potential adoption by early-stage legal professionals.

With a significant portion of the participants stating that there will be a need for AI skills in the legal domain now (36%) and in 5 years (51%) the attention rightfully falls on what can be done to expand the skillset of those already in the field or those who are law students (considering that discussions mostly center around the future, but what about the now?). What individuals can do is self-educate on understanding AI and machine learning basics, data science, programming skills, prompt engineering and try to gain practical experience with AI systems and legal AI systems. Online courses such as those offered on websites like Coursera[34] (on topics like LLMs and prompt-engineering) could be beneficial. The overlap between the Bloomberg best current legal AI Tool options referenced in the Introduction and the tools our survey participants use consists of only HarveyAI and ChatGPT. This probably reflects the fact that the majority of the participants are 1st year law students (36%), followed by 3rd year students (21%) and LL.M students (26%) yet not fully exposed to these tools especially considering that most of them require subscriptions by institutions. Another important aspect of self-education focuses on ethics, security and privacy of AI in all aspects of the legal field to ensure the appropriate use of the technology including for example scenarios such as the legal ramifications of a company sharing employee data without consent with a company that builds LLMs. Familiarity with AI interfaces and AI terminology would also prove beneficial as it would provide a practical and easily applicable foundation for those seeking to expand their knowledge. For those aiming to gain more in-depth knowledge, joint degrees between law and AI are available at some universities.

Our study participants are very clear in their views on the applications of AI in the legal field – 73% of the respondents state using AI tools for help with school and work (together and respectively) with the main areas of usage being text generation, summarization, legal research, information extraction, data science and planning. Notably, very few utilize AI tools for nuance-requiring tasks (e.g. arbitration) or accuracy-related tasks (e.g. referencing), and specific legal domain tasks. This likely reflects the early stage of the participants' career development but it also highlights the limitations of the technology and provides valuable insights on the need for further improvement and solutions for the full integration of AI in the legal field aligning with the concerns expressed in Ma et al[5]. Accuracy and privacy appear as the primary challenges for the broader adoption and integration of AI in the legal field especially for specialized legal functions. Addressing the issue of accuracy, from a technology development point of view combining LLMs with RAG systems could be the key solution to improving the accuracy and relevance of LLM outputs by removing the need for additional training[5]. The RAG system will provide real-time access to specific documents and data points which in turn will help with the reduction of hallucinations and the improvement of the adaptability, diversity, explainability and efficiency of the LLM. Open models (e.g. Llama) would provide a better alternative for those looking for enhanced privacy by providing more transparency in data processing and handling which in turn satisfies the growing demand for security within the legal field (and any high-stakes field for that matter).

The technical complexities of the AI field are reflected in the current state of AI regulations and the awareness of these regulations among law students. The data reveal that the majority of the participants (63%) are not keeping up with AI regulations with the EU AI Act[17] being the most followed. Additionally, 40% of the participants are unsure of whether regulations are keeping up with innovation which shows the uncertainty among law students and the gaps to address. There is a general uncertainty in how, when, where and how much AI usage is allowed and acceptable within the legal field. Governments, employers and educational institutions need to set clear guidelines and policies to make sure ambiguity is removed and keep the AI usage process fair and ethical. A recent case[35] of the US-based law firm Levidow, Levidow & Oberman exemplifies the dangers of naive AI usage[36]. Two of their lawyers used non-existent citations in an aviation injury claim generated by ChatGPT believing the technology to be a search engine and not aware of its tendency to hallucinate. A potential solution would be to build an international united codex, applicable in every jurisdiction which clearly states the rules around the usage of AI systems and ensure compliance.

There is a prevailing sense of concern and caution about the future of AI in the legal field. Our study participants are concerned that AI might affect productivity and lead to intellectual laziness and overreliance on the technology as well as a decline in job satisfaction. Moreover, the concern reflects the rapid integration of AI and its potential to disrupt traditional legal practices. However, the study participants remain hopeful about the future, if the technology gets mindfully deployed without displacing human capital with the focus on streamlining mundane and time-consuming tasks in the human-in-the-loop scenario. Furthermore, this cautious attitude offers insights for educational institutions and companies to act in advance and prepare the workforce for the evolving AI landscape through formal training (discussed above) and continued education. The future of the legal field, to a degree, depends on fostering a balance between innovation and preservation of human expertise.

Speaking of balance, it is important to mention that AI in the legal field follows the pattern of disruptive technologies. The Innovator's Dilemma by Clayton M. Christensen[37] encapsulates the essence of this ideology by explaining why most firms (or any institution for that matter) miss out on new waves of innovation. No matter the industry, he says, a successful company with established products will get pushed aside unless it knows how and when to abandon traditional business practices. Christensen also leans into the pace of technological advancement and its tendency to outstrip needs. Viewing AI in the legal field as a disruptive technology, we all falter in how to approach it as it falls outside of our established patterns. This likely explains the cautious attitude of our study participants – besides handling the load of their law studies, they need to figure out what this new technology is all about and how that would affect their careers. It should also be noted that Christensen's model indicates that existing successful companies, when faced with a disruptive technology, in general only survive by reinventing within; usually through establishing a new internal group largely isolated from and unencumbered by current business processes to grow and exploit the new technology. Overtime, this group typically grows and ultimately consumes and replaces the old company. If this model holds for the legal domain and AI, the opportunities for law students with training in AI and largely "uncontaminated" by the existing legal models may be significant.

With an eye on the future, our plan is to initiate an AI and Commercial Awareness Law Student Club. With 59% of the participants interested, the club is swiftly gaining popularity. The aim of the club is to provide a venue to keep up with the developments of the AI and legal domain and potentially initiate collaborations with Computer Science Departments on applications of AI in the legal field. Another potential idea is to open a line of communication with university administrations on creating AI courses for law students to prepare law students for the skills needed for the world of the future. Additionally, the Club intends to regularly discuss the evolving AI skillset with the goal of spreading awareness of the changing skillset needed for a successful legal career in pace with AI developments.

Although this paper presents the first more in-depth study on the topic of law students' perspective of AI in the legal field and includes more variables than previous work, it is not without limitations. Our survey includes 27 questions answerable in about 20 minutes cognizant of time limitations by the participants. Although the questions aim to cover the most important variables, an expanded set of variables would be even more elucidating. In addition, the majority of the survey participants ended up being from the UK. As future work, we would like to expand the study to include a larger cohort more evenly distributed across locales. In addition, it would be insightful to repeat the study with the current cohort in one year's time to track how the students' viewpoints evolve.

## Conclusion

While there are studies outlining the landscape of AI in the legal field as well as surveys of the current AI efforts of law firms, to our knowledge there has not been an investigation of the intersection of law students and AI. Such research is critical as the current law students are the future legal professionals and a well-prepared workforce translates into not only operational efficiencies for the firms but also into adequately prepared next generation of jurists who will regulate AI as well as represent and adjudicate AI-related cases. This paper presented, to our knowledge, the first study of this intersection of law students and AI. The results of the study could serve as the basis to bridge the gap between AI knowledge and training on one hand, and AI applications in the legal field on the other. Bridging the gap could take many forms: educational institutions including AI training, governments working on the issue of misinformation and building a universal AI codex, employers providing training and information on AI to their employees from established experts, and individuals self-learning and acquiring the AI skills needed for a modern legal career. Importantly, the results of this study demonstrate the uniqueness of law students as a distinct cohort when researching AI in the legal field and the need for in-depth investigations to understand law students' AI awareness and concerns.

## Acknowledgments

We are very grateful to the participants in the survey who enthusiastically volunteered their time. We thank Prof. Sinha, London Metropolitan University.

# Appendix

*Q1: Type of student*

| | |
|---|---|
| LLB student | 63% |
| MS | 22% |
| Other (e.g. working at a law firm, lawyer) | 9% |
| BA law student | 6% |
| TOTAL | 100% |

*Q2: Study year (for participants who indicated)*

| | |
|---|---|
| 1st year | 36% |
| Master's studies | 26% |
| 3rd year | 21% |
| 2nd year | 15% |
| 4th year | 2% |
| TOTAL | 100% |

*Q3: University name (for the participants who indicated)*

| | |
|---|---|
| London Metropolitan University | 29% |
| ULaw | 11% |
| BPP University | 8% |
| University of London | 5% |
| University of Warwick | 5% |
| Àjàyí Crowther University | 3% |
| Aston University | 3% |
| BCU | 3% |
| College of Legal Practice | 3% |
| Hamdard School of Law, Pakistan | 3% |
| National Law University, Delhi | 3% |
| Stirling Univeristy | 3% |
| UCFB | 3% |
| University of Essex | 3% |
| University of Exeter | 3% |
| University of Linclon | 3% |
| University of Manchester | 3% |
| University of Plymouth | 3% |
| University of Reading | 3% |
| University of Southampton | 3% |
| TOTAL | 100% |

*Q4: Understanding of AI*

| | |
|---|---|
| Only as a user | 43% |
| Solid but non-technical | 41% |
| Solid with technical details | 13% |
| In depth | 2% |
| No understanding at all | 2% |
| TOTAL | 100% |

*Q5: Formal AI Training*

| | |
|---|---|
| No | 91% |
| Yes | 9% |
| TOTAL | 100% |

*Q6: Usage of AI applications*

| | |
|---|---|
| Yes | 70% |
| No | 30% |
| TOTAL | 100% |

*Q7: Length of usage of AI applications*

| | |
|---|---|
| More than a month but less than 6 months | 28% |
| More than a year but less than 2 years | 28% |
| More than 6 months but less than a year | 20% |
| Less than a month | 15% |
| 2+ years | 10% |
| TOTAL | 100% |

*Q8: Frequency of usage of AI applications*

| | |
|---|---|
| Weekly, less than 5 times | 38% |
| Daily multiple times | 28% |
| Weekly, more than 5 times | 18% |
| Other (ad-hoc basis, very rarely) | 10% |
| Daily, less than 3 times | 8% |
| TOTAL | 100% |

*Q9: Types of AI applications used (for respondents with Yes for Q6)*

| | |
|---|---|
| ChatGPT 3.5 (free version) | 65% |
| ChatGPT 3.5 (free version) and ChatGPT 4.0 (subscription) | 20% |
| ChatGPT 4.0 (subscription) | 15% |
| TOTAL | 100% |

*Q10: Purpose of AI usage (for respondents with Yes to Q6: multiple answers)*

| | |
|---|---|
| To help with school and work | 40% |
| To help with school | 17% |
| To help with work | 17% |
| Planning and scheduling | 6% |
| Writing emails | 4% |
| Cooking recipes | 2% |
| Financial investment | 2% |
| Job applications | 2% |
| Other creative engagements | 2% |
| Technical theory questions | 2% |
| Thesaurus | 2% |
| To get a summary | 2% |
| To understand complex topics | 2% |
| TOTAL | 100% |

*Q11: Helpfulness of AI (for respondents with Yes to Q6)*

| | |
|---|---|
| Extremely helpful | 37% |
| Somewhat helpful | 34% |
| Helpful | 27% |
| Not so helpful | 2% |
| TOTAL | 100% |

***Q12: Prejudice/bias against AI***
No                                                                              80%
Yes                                                                             20%
TOTAL                                                                          100%

***Q13: The importance of AI in the legal field***
Extremely important                                                             39%
Important                                                                       35%
Somewhat important                                                              26%
TOTAL                                                                          100%

***Q14: Attitude towards AI in the legal field***
Cautiously hopeful                                                              44%
Mostly hopeful                                                                  30%
Excited                                                                         20%
Extremely concerned/worried                                                      6%
TOTAL                                                                          100%

***Q15: Usage of AI applications in the legal field***
No                                                                              80%
Yes (Aria, Chatsonic, Harvey, ContractPodAi, Luminance, ChatGPT)                20%
TOTAL                                                                          100%

***Q16: Inclusion of AI in the legal ciriculum (multiple answers)***
Very important                                                                  40%
Somewhat important                                                              26%
Important                                                                       21%
Other                                                                            7%
My university already includes such training                                     3%
Not important                                                                    3%
TOTAL                                                                          100%

***Q17: Collaboration with Computer Science department on applications in the legal field***
Very enthusiastic                                                               37%
Enthusiastic                                                                    37%
Somewhat enthusiastic                                                           17%
Not interested                                                                   9%
TOTAL                                                                          100%

***Q18: Level of preparation for the future usage of AI in the legal field***
Somewhat prepared                                                               46%
Not prepared                                                                    31%
Well prepared                                                                   11%
Excellently prepared                                                             9%
N/A                                                                              2%
TOTAL                                                                          100%

*Q19: Need for AI skills and background in the future of the legal domain*

| | |
|---|---|
| In 5 years | 51% |
| Now | 36% |
| In 10 years | 8% |
| Never | 6% |
| TOTAL | 100% |

*Q20: Your opinion on impactful applications of AI in the legal field (multiple answers)*

| | |
|---|---|
| Drafting - contracts | 20% |
| Summarization | 17% |
| Legal research | 10% |
| Information extraction | 8% |
| Data science - collection, extraction, tracking, data-driven decision making | 7% |
| Drafting - other, pleadings, judgements | 7% |
| All aspects | 6% |
| No idea/not sure | 2% |
| Application forms | 1% |
| Arbitration | 1% |
| Automation - contracts | 1% |
| Case finding | 1% |
| Contract law | 1% |
| Copyright | 1% |
| Corporate emails | 1% |
| Corporate law | 1% |
| Criminal law | 1% |
| Drafting - statistics | 1% |
| Initial triage | 1% |
| Interpretation of judgements, precedents, etc. | 1% |
| Judicial administration | 1% |
| Marriage/Deeds of assignment | 1% |
| Mundane areas | 1% |
| Topic understanding | 1% |
| Visa generation | 1% |
| TOTAL | 100% |

*Q21: Keeping up with the landscape of AI regulations*

| | |
|---|---|
| Not really | 63% |
| Somewhat | 31% |
| Yes | 6% |
| TOTAL | 100% |

*Q22: Types of regulations followed (for respondents with Yes to Q21, multiple answers)*

| | |
|---|---|
| The EU AI Act | 60% |
| The Office of Artificial Intelligence | 40% |
| TOTAL | 100% |

*Q23: Adequacy of the speed of innovation vs regulation catch up*

| | |
|---|---|
| Not sure | 40% |
| No | 38% |
| Yes | 23% |
| TOTAL | 100% |

*Q24: Following AI cases e.g. NewYork Times v Open AI*

| | |
|---|---|
| No | 81% |
| Yes | 19% |
| TOTAL | 100% |

*Q25: AI's influence on opinions*

| | |
|---|---|
| Yes | 65% |
| No | 31% |
| Not sure | 2% |
| Somewhat | 2% |
| TOTAL | 100% |

*Q26: Interest in AI and Commercial Awareness Law Student Club*

| | |
|---|---|
| Yes | 59% |
| No | 41% |
| TOTAL | 100% |

*Table 1: Topics of survey questions and answer distribution, percentages are rounded. Note we list the topic of the question, not the verbatim question. The sizes of the colored bars denote frequency, e.g. a longer bar indicates bigger percentage. Background questions in blue. AI Usage questions in red. AI Applications in the Legal Field in green. AI Regulations in yellow. Conclusion questions in purple.*